\documentclass[useAMS,usenatbib]{mn2e}

\usepackage{amsmath,amssymb,graphicx}

\newcommand{\kepler}{\textit{Kepler}}

\title[Eclipse Timing Variations in KOI 928]{The architecture of the hierarchical triple star KOI 928 from eclipse timing variations seen in \kepler\ photometry}
\author[J. H. Steffen \textit{et al.}]{J. H. Steffen$^{1}$\thanks{E-mail: jsteffen@fnal.gov},
S. N. Quinn$^2$, 
W. J. Borucki$^3$,
E. Brugamyer$^4$,
S. T. Bryson$^3$, \newauthor
L. A. Buchhave$^{5}$
W. D. Cochran$^4$,
M. Endl$^4$, 
D. C. Fabrycky$^6$, 
E. B. Ford$^7$, \newauthor
M. J. Holman$^2$,
J. Jenkins$^{3,8}$,
D. Koch$^3$,
D. W. Latham$^2$,
P. MacQueen$^4$, \newauthor
F. Mullally$^{3,8}$,
A. Pr\v{s}a$^{9}$,
D. Ragozzine$^2$, 
J. F. Rowe$^3$,
D. T. Sanderfer$^3$, \newauthor
S. E. Seader$^{3,8}$
D. Short$^{10}$,
A. Shporer$^{11}$,
S. E. Thompson$^{3,8}$,
G. Torres$^2$, \newauthor
J. D. Twicken$^{3,8}$
W. F. Welsh$^{10}$,
G. Windmiller$^{10}$\\
$^1$Fermilab Center for Particle Astrophysics, P.O. Box 500, Batavia, IL 60510, USA\\
$^2$Harvard-Smithsonian Center for Astrophysics, 60 Garden St., Cambridge, MA 02138, USA\\
$^3$NASA Ames Research Center, Moffett Field, CA 94035, USA\\
$^4$McDonald Observatory, The University of Texas, Austin, TX 78712-2059, USA\\
$^5$Niels Bohr Institute, Copenhagen University, DK-2100 Copenhagen, Denmark\\
$^6$Department of Astronomy and Astrophysics, University of California, Santa Cruz, Santa Cruz, CA 95064, USA\\
$^7$Department of Astronomy, University of Florida, 211 Bryant Space Science Center, Gainesville, FL 32611-2055, USA\\
$^{8}$SETI Institute, 515 North Whisman Road, Mountain View, CA, 94043\\
$^{9}$Villanova University, Department of Astronomy and Astrophysics, 800 E. Lancaster Ave., Villanova, PA 19085\\
$^{10}$San Diego State University, 5500 Campanile Drive, San Diego, CA 92182, USA\\
$^{11}$Las Cumbres Observatory Global Telescope, Goleta, CA 93117, USA
}
\begin{document}

\bibliographystyle{plainnat}

\maketitle

\label{firstpage}

\begin{abstract}
We present a hierarchical triple star system (KIC 9140402) where a low mass eclipsing binary orbits a more massive third star.  The orbital period of the binary (4.98829 Days) is determined by the eclipse times seen in photometry from NASA's \kepler\ spacecraft.  The periodically changing tidal field, due to the eccentric orbit of the binary about the tertiary, causes a change in the orbital period of the binary.  The resulting eclipse timing variations provide insight into the dynamics and architecture of this system and allow the inference of the total mass of the binary ($0.424 \pm 0.017 \text{M}_\odot$) and the orbital parameters of the binary about the central star.
\end{abstract}

\begin{keywords}
Eclipsing Binaries, \kepler, KIC 9140402, FERMILAB-PUB-11-292-AE
\end{keywords}

\section{Introduction}

The timings of transit or eclipse events in multibody astronomical systems provide a high precision measurement of the phase of the orbits of the transiting bodies---typically a few parts in 10$^4$ or better.  Such high precision measurements allow for detailed studies of the dynamics of these systems through eclipse timing variations or ETVs (or transit timing variations, TTVs, for planetary systems) \citep{Borkovits:2003,Agol:2005,Holman:2005}.  A variety of mechanisms can cause the eclipse times to deviate from a linear ephemeris including the R\o mer effect (light travel time, or LTT), transverse displacements of the star with respect to the system barycentre, resonance interactions among the bodies, and effects that correspond to the synodic periods of the objects.  A detailed discussion of these cases is found in \citet{Agol:2005}.

One notable cause of ETVs is the effect of a changing tidal field on a binary pair due to a perturber on a hierarchical, eccentric orbit.  This scenario was derived analytically in \cite{Borkovits:2003} and a simplified derivation is shown in \cite{Agol:2005}.  Basically, when the perturbing third object is far from its short-period binary companions, the period of the binary is largely unmodified.  However, when the perturbing object is near the binary and near the binary's orbital plane, its presence slows the orbital period of the binary.  The result is a periodic TTV signal with a period equal to the orbit time of the perturbing body.  The more eccentric the orbit, the larger and more asymmetric the TTV signal appears because the slowing of the binary's orbital period at the perturber's pericentre passage takes a smaller fraction of its orbital period and the large change in proximity from the high eccentricity exaggerates the change in the binary's period.  The nature of this signal is such that it is virtually independent of the azimuthal orientation of the apse of the orbit of the perturbing body, though it does depend upon the mutual inclination of the two relevant orbital planes.

The first system known to exhibit this effect, also found with \kepler\ photometric data, is the triple star system KOI 646 \citep[KIC 5384802, ][]{Fabrycky:2010}.  Here we present and discuss a second stellar system that shows a periodic ETV signal consistent with this same model, KOI 928 (KIC 9140402).  A third \kepler\ system that shows similar orbital architecture, but that is viewed in a different orientation---and thus does not show the same ETV signal, is KOI 126 \citep[KIC 5897826, ][]{Carter:2011}.  Additional star systems that show trends indicative of dynamical interactions were reported in \citet{Slawson:2011}.  The paper is organized as follows.  In Section \ref{photometry} we present the \kepler\ photometry and transit times.  In Section \ref{spectroscopy} we outline the spectroscopically derived stellar parameters and radial velocity (RV) measurements of the target star.  Section \ref{analysis} gives the dynamical analysis of the transit times and radial velocity measurements of the system.  Concluding remarks are in Section \ref{discussion}.  We note that the true orbital structure of these systems is a bright central star orbited by an eclipsing binary of low mass stars.  However, for the purposes of our discussion we will label the eclipsing binary as the ``inner binary'' of objects one and two and the third star, which perturbs the orbital period of the inner binary, as the third or ``outer'' object.

\section{\kepler\ Photometry}\label{photometry}

\begin{figure}
\includegraphics[width=0.45\textwidth]{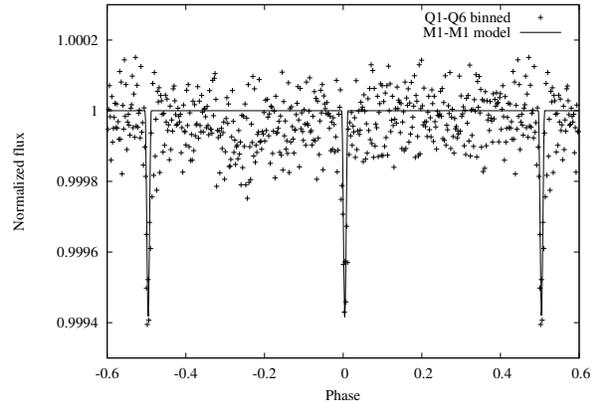}
\caption{Binned and folded light curve for KOI 928.  Also shown is a representative model which has equal size ($R = 0.238 R_\odot$) and mass ($M = 0.21 M_\odot$) members of the binary.\label{timeseries}
}
\end{figure}

KOI 928 (KIC 9140402) has \kepler\ magnitude Kp = 15.251, making it quite dim among \kepler\ targets.  It is located at RA 18:59:02.26 and Dec 45:35:56.86.  For our study, we use data from the first six quarters of \kepler\ operations (BJD 2454968 -- 2455650) corresponding to nearly 700 days of observation.  Information about the \kepler\ spacecraft and its performance can be found in \citet{Koch:2010}.  The period of the eclipse events is 4.98829 Days and the eclipse depths are 0.06\% of the nominal flux.  A binned lightcurve and representative model (generated using the PHOEBE software from  \citet{prsa:2005}), is shown in Figure \ref{timeseries}.

This system was initially identified as a planetary candidate through the Transiting Planet Search and the Data Validation Pipelines \citep{Jenkins:2010,Wu:2010} which identify significant transit-like features and conduct a battery of statistical tests on those transit events in an effort to rule out false-positive transit signals.  As data for KOI 928 were being analyzed, the interpretation of the system quickly grew complicated.  The transit times showed a sizeable, roughly sinusoidal timing variations with nearly a two-hour peak-to-peak amplitude (consistent with a near-resonant two-planet system).  However, intial RV measuremen nts (described in the next section) differed significantly from the predictions of a two-planet model.

A PSF fit to the difference image formed by subtracting averaged in-transit pixels from averaged out-of-transit pixels \citep[see][]{Torres:2011} indicated no significant centroid motion.  This fact effectively eliminates the possibility of the transits being on a star that is more distant than 0.3 pixels (1.2'') from KOI-928.  Taken together, with a few additional RV data, the information suggests a model of a bright star being orbited by a low-mass binary pair.

Attempts to model the eclipse times of this system suffer both from the lack of photons (given the dimness of the target) as well as additional systematic errors.  In particular, some estimates of the eclipse times show multiple local minima while others have unusually large error bars.  Consequently, we derived the eclipse times and their uncertainties using two different methods (described below).  We then adopt one set of times as ``nominal'', but eliminate eclipses at certain epochs based upon the estimated errors and the differences between the two methods.  Times from the second method are not analyzed in the dynamical model.

To estimate the nominal eclipse times for both the members of the eclipsing binary, we fit standard 4$^{\rm th}$ order non-linear limb-darkened eclipse models \citep{Mandel:2002} to the Kepler light curve.  For each primary or secondary eclipse, the model allowed for independent values of the primary-secondary radius ratio, eclipse duration, and impact parameter.  For each eclipse, we fit for the flux normalization and a local linear slope in the flux and we numerically average the model over the 30 minute integration duration.

The first step to determine the eclipse times is to fit a single model to the set of all eclipses for each individual star assuming a constant orbital period.  Second, we hold the radius ratio, eclipse duration, and impact parameter fixed, and fit a small segment of the light curve around each eclipse for the remaining parameters.  Third, we phase the light curve using each measured eclipse and refit for the eclipse parameters (aside from period and epoch).  The second and third steps are iterated to converge on a final model.  This model also includes a nuisance parameter which estimates the ``contamination'' light from stars in the Kepler aperture (excluding the star being eclipsed) as a fixed parameter.  We verified that our results for eclipse times are not sensitive to the value of this nuisance parameter.

The second, comparative set of eclipse time estimates were calculated using an iterative process starting with an initial linear ephemeris and eclipse width.  Eclipses in the ``raw'' data were masked, then the light curve was piecewise detrended and normalized locally (0.5 days) using a cubic polynomial.  All the eclipses were then folded on the ephemeris, and a piecewise cubic Hermite spline was fitted using least-squares to the folded eclipse using observations that fell within a window of width $1.4\times$ the eclipse width + 1 cadence on each side.  The cubic Hermite spline was fit using 9 evenly spaced points across the window and the $\chi^{2}$ of the fit recorded.  The cubic Hermite was refit 25 more times using odd numbers (11-35) of spline knots.  The best fit from these 26 cases defined the eclipse template, and the eclipse width estimate was updated using this template.

The light curve was again detrended locally using a cubic polynomial but now using three different out-of-eclipse lengths: 7.5 hours, 15 hours, and 20 hours.  For each of the three out-of-eclipse lengths, the template was correlated with the eclipse at 1000 time steps, spanning 115 min.  The estimate for the mid-eclipse time is that which gave the minimum $\chi^{2}$ value over the three different out-of-eclipse lengths.

This entire process was iterated using the new eclipse time estimates, but those eclipses with a reduced $\chi^{2} > 2.0$ were eliminated from the template building step, and the correlation time step length was reduced by a factor of 8. Once the second iteration was completed, the uncertainties in eclipse times were estimated using the $\chi^{2}$ curve of the fits.

To select the eclipse times from the nominal (first) method to  used in our analysis, we rejected those epochs where either of the two methods had large uncertainties (greater than 0.045 days) and the epochs where the two methods disagree by more than 3$\sigma$.  The epochs that survive these cuts were analyzed.  These two criteria were determined by analyzing the distribution of the differences in the eclipse times of the two models and the distribution of the uncertainties in the eclipse times.  In both cases (the cut on uncertainty and the cut on difference) there is an obvious gap where the outlier population dominates over the nominal distribution and the chosen cuts reflect those transitions.  The eclipse times used for this analysis are given in the appendix.

\section{Stellar Properties, Imaging, and Spectroscopic Observations} \label{spectroscopy}

We obtained seven high-resolution spectra of KOI 928 in order to measure improved stellar properties of the bright, outer star and to place radial velocity constraints on its orbit.  Six spectra were taken with the Tull Coud\'e Spectrograph on the 2.7m Harlan J. Smith Telescope at the McDonald Observatory in west Texas, which has a resolving power of $R\approx60,000$ and wavelength coverage $3750-10000$ \AA.  One additional spectrum was taken with the Fiber-fed Echelle Spectrograph (FIES) on the 2.5m Nordic Optical Telescope (NOT) at La Palma, Spain \citep{Djupvik:2010}.  The FIES spectrum was taken with the medium-resolution fiber, which has resolving power of $R\approx46,000$ and wavelength coverage $3600-7400$ \AA.

In order to determine the effective temperature ($T_{\text{eff}}$), projected rotational velocity ($v \sin{i}$), surface gravity ($\log g$), and metallicity ([Fe/H]) of the bright star in the system, we cross-correlated the strongest spectrum---the only one with signal-to-noise ratio (SNR) greater than 20 per resolution element---against a grid of synthetic stellar spectra computed from Kurucz models \citep{Kurucz:1992}.  A new set of tools \citep{Buchhave:2011} was then used to derive more precise stellar parameters from the normalized cross correlation peaks.  Formally, the value of $\log g = 4.56$ places the star below the isochrones in an unphysical part of the H-R diagram.  This is most likely due to errors in the measured quantities and given the relatively low SNR of our spectrum and the strong spectroscopic correlations between $T_{\text{eff}}$, $\log g$, and [Fe/H], this is not surprising.  However, the formal error in $\log g$ is large enough that there are valid solutions that do fall on the isochrones.  The results from this analysis, with conservative uncertainties, are given in Table \ref{stellarparams}.

To obtain radial velocities, we performed a multi-order cross-correlation of the six McDonald spectra following the procedure outlined in \citet{Buchhave:2010}.  For the FIES spectrum, we adopt the RV derived from cross-correlation against the best-matched synthetic template.  The velocities are shifted onto the IAU absolute scale as defined by the velocity of the IAU RV standard HD 182488 \citep{Nidiver:2002}.  The errors have been inflated to include an instrumental component corresponding to the long term RMS velocity residuals of HD 182488 as observed by each instrument.  The RV measurements derived from this analysis are given in Table \ref{rvtable}.

\begin{table}
\begin{center}
\caption{Stellar parameters for the central star in KOI 928.\label{stellarparams}}
\begin{tabular}{lrr}\hline \hline
Parameter & Value & Uncertainty \\ \hline
$T_{\text{eff}} (K)$ & 5506 & 150 \\
$\log g$ & 4.56 & 0.23 \\
$[Fe/H]$ & 0.08 & 0.29 \\
Vsini (km/s) & 3.3 & 1.7 \\
$M_3$ ($M_{\odot}$) & 0.97  & 0.1 \\
$R_3$ ($R_{\odot}$) & 0.89  & 0.1 \\ \hline
\end{tabular}
\end{center}
\end{table}

\begin{table}
\caption{Radial velocity measurements.\label{rvtable}}
\begin{tabular}{rrrr} \hline \hline
Date & Radial Velocity & Error & Instrument \\
(BJD - 2454900) & (km/s) & (km/s) & \\ \hline
445.8325 & -8.039 & 0.339 & MCD \\
523.5395 & 10.322 & 0.400 & FIES \\
570.6033 &-14.269 & 0.319 & MCD \\
596.6075 & -4.859 & 0.343 & MCD \\
599.6062 & -2.119 & 0.340 & MCD \\
627.5932 & 11.821 & 0.319 & MCD \\
732.9574 &  9.697 & 0.520 & MCD \\ \hline
\end{tabular}
\end{table}

\section{Dynamical Model}\label{analysis}

As discussed above, the case of a hierarchical triple system where the distant third body is on an eccentric orbit, the changing tidal field produced by the perturbing third body causes a change in the period of the binary that cycles with its orbit about the perturber.  For our investigation, we use the coplanar approximation as model fits with mutually inclined orbits did not produce a sufficent improvement to justify the additional parameters.  Thus, our model is given by (equation (25) in \cite{Agol:2005}):
\begin{equation}\label{modeleq}
\delta t = \xi \frac{P_3}{(1-e_3^2)^{3/2}} \left[ f_3 - \frac{2 \pi (t-\tau_3)}{P_3} + e_3 \sin f_3 \right]
\end{equation}
where
\begin{equation}
\xi \equiv \frac{1}{2\pi} \left( \frac{m_3}{m_1+m_2} \right)  \left( \frac{P_{12}}{P_3} \right)^2
\end{equation}
and where $m_1$ and $m_2$ are the masses of the two objects in the binary, $P_{12}$ is the period of the binary.  The parameters for the third body are its mass $m_3$, period $P_3$, eccentricity $e_3$, time of pericentre passage $\tau_3$, and true anomaly $f_3$ (we change notation from \cite{Agol:2005} to use the ``3'' subscript to denote the third body).  The ETV effect for this coplanar case is independent of the orientation of the orbit of the third body with respect to the observer (i.e., the longitude of pericentre $\varpi_3$).  This orientation can be measured through the LTT effect and with the RV data.  For KOI 928, the timing uncertainties are too large to provide meaningful contraints from LTT alone and require the inclusion of RV measurements in the analysis to identify the value of this parameter.

The seven model parameters for our analysis include the mass ratio $M \equiv m_3/(m_1+m_2)$, $P_{12}$, $P_3$, $\tau_3$, $e_3$, $\varpi_3$, and the ephemeris epoch $T_0$ in $\text{BJD}-2454900$.  The mass ratio is not well constrained by the transit data without an estimate for one of the masses (either the mass of the perturber or the mass of the binary).  Consequently, we fix the mass of the pertuber to the value determined from the spectroscopy.

The RV data and timing data were fit to the ETV model in equation \ref{modeleq}, with additional terms for the geometric LTT effect and the RV signal, using a Markov Chain Monte Carlo (MCMC).  We assume measurement uncertainties are Gaussian and uncorrelated.  The model parameters corresponding to the maximum likelihood model and the 68.3\% credible intervals are given in Table \ref{modparams}, as well as the median value among a posterior sample for each model parameter.  (Note that this set of median values does not
correspond to any specific model.)

\begin{figure}
\includegraphics[width=0.45\textwidth]{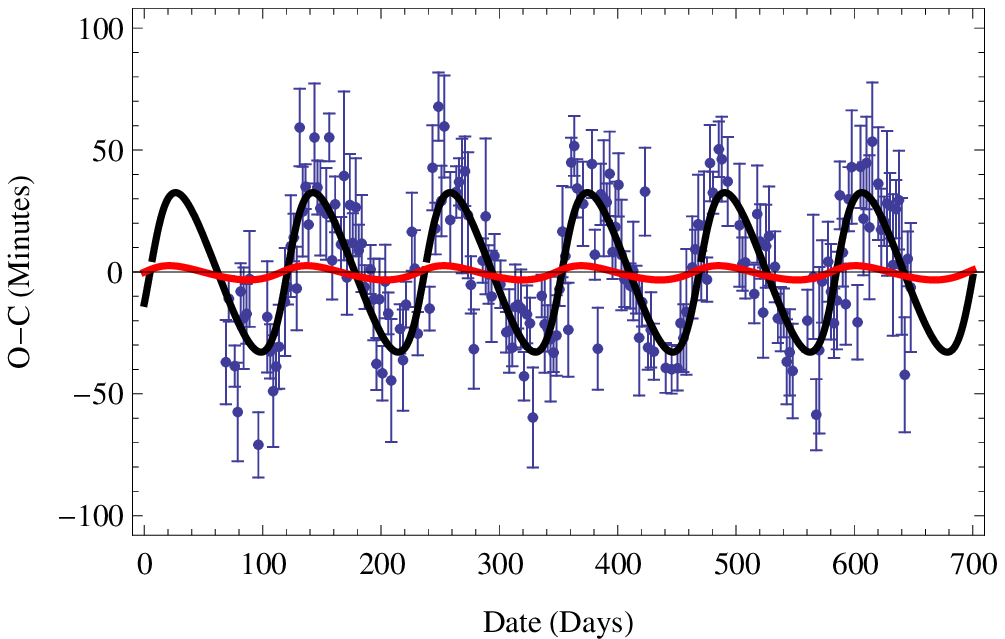}
\includegraphics[width=0.45\textwidth]{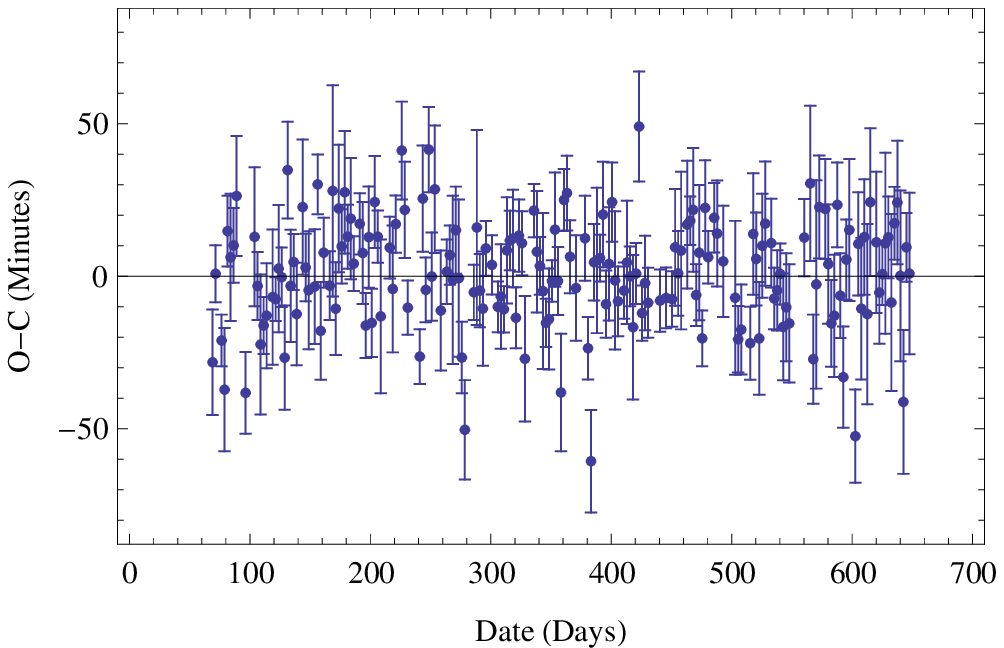}
\caption{(Top) Plot of the timing data and the best fitting model.  A second (barely visible) red curve shows the LTT effect which is the only means to measure the orbital orientation $\varpi_3$ without RV measurements.  (Bottom) Residuals after subtracting the model transit times.}
\label{solutionttv}
\end{figure}

\begin{figure}
\includegraphics[width=0.45\textwidth]{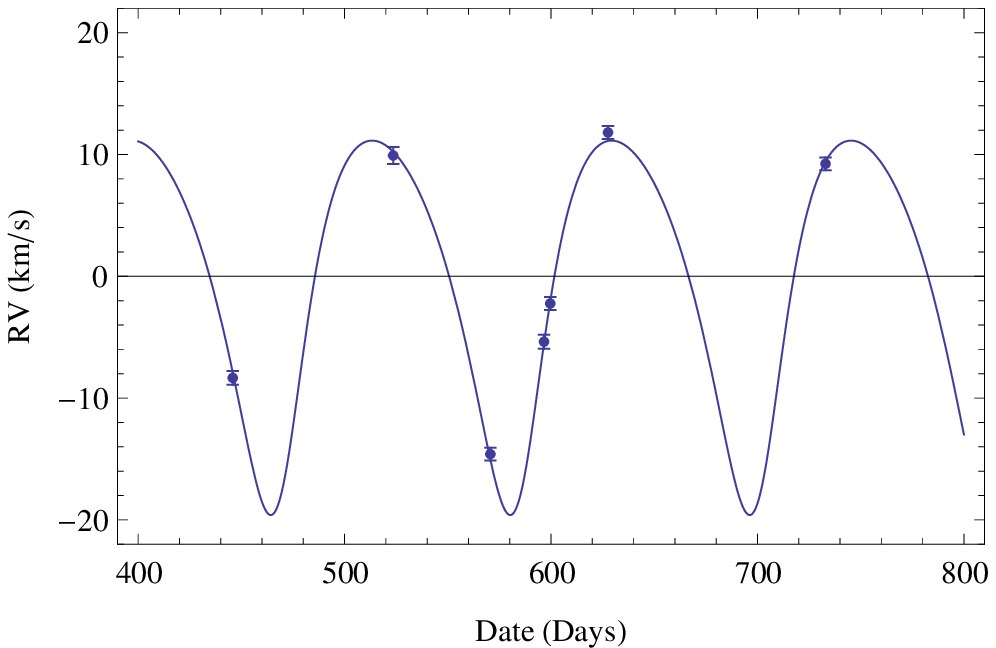}
\includegraphics[width=0.45\textwidth]{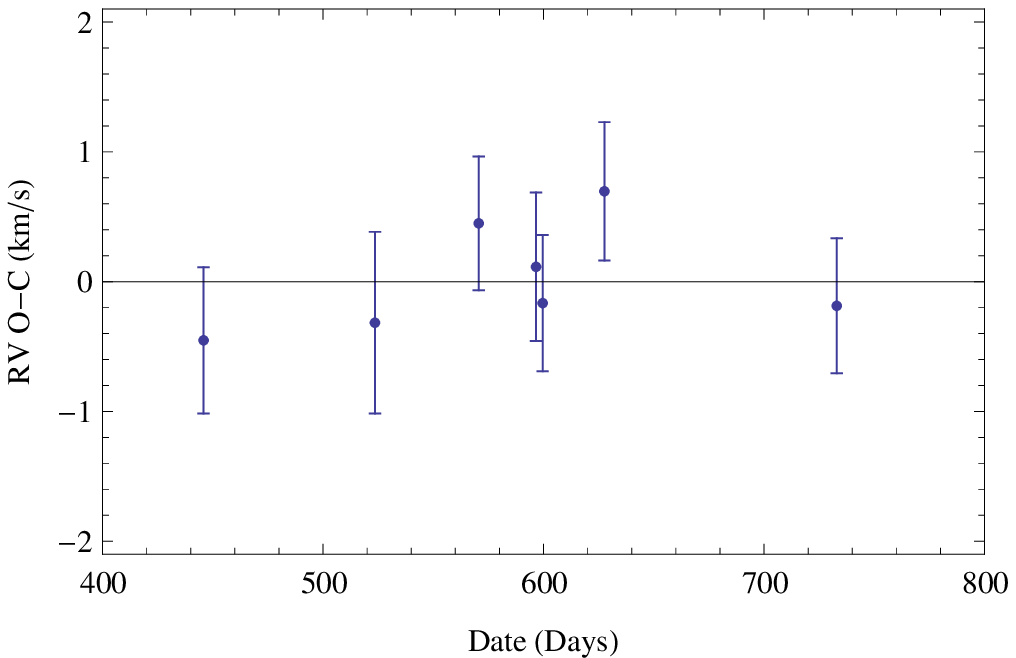}
\caption{(Top) Plot of the RV data and the best fitting model.  (Bottom) Residuals after subtracting the model transit times.}
\label{solutionrv}
\end{figure}

The uncertainties in the model parameters are found using the corresponding posterior distributions from the MCMC.  After rejecting the first $\sim 20\%$ of the chain, the values for each of the model parameters were sorted and the smallest and largest 15.9\% of the values were rejected.  The mean difference between the median of the remaining values for each parameter and the largest and smallest values is our estimate for the uncertainty in that parameter.  There is some small asymmetry in the distributions, but it has a sufficiently small effect that we do not report two-sided error bars.  In addition, we study the autocorrelation of the chains in the model parameters to determine the uncertainty in our error estimates.  The correlation lengths of each parameter indicates a worst-case uncertainty of 10\% in the error estimate, while several parameters are much better\footnote{The relevant quantity being the number of correlation lengths in a Markov chain rather than the number of links in the chain.}.  The best fitting model and the residuals are shown in Figures \ref{solutionttv} and \ref{solutionrv} for the eclipse times and RV measurements respectively.

Given the parameter values obtained from the dynamical model, an additional analysis was conducted on the light curve in order to determine the sizes of the stars in the binary.  Given that the eclipse depths are almost indistinguishable, we assumed that the two stars are identical in size and mass (with masses equal to 0.212\,M$_\odot$).  The result of this analysis (also shown in Table \ref{modparams}) is that the radii of the two binary members are $0.28 \pm 0.05 \text{R}_\odot$.  These radii estimates are somewhat larger than the isochrone models for low-mass stars given in \citet{Baraffe:1998}, however, the discrepancy is not significant.  Moreover, the rather large uncertainty in these sizes makes them less useful for comparison to other measured systems.  Additional photometric data from \kepler\ should lessen the uncertainty in this parameter and consequently provide more valuable insight into the physical properties of such stars and our modelling of them.

\begin{table}
\begin{center}
\caption{Parameter values for the KOI 928 system}
\protect\label{modparams}
\begin{tabular}{cccc} \hline \hline
Parameter & Value & Error & Best Fitting    \\ \hline
$M_3$ & 0.97 ($M_\odot$)       & 0.1* &     0.97 \\
$M_{12}$  & 0.424 ($M_\odot$)  & 0.017$^\dag$ &    0.423808\\
$P_{12}$ & 4.988287 (Days)     & 0.000015 & 4.988284 \\
$P_3$ & 116.03 (Days)          & 0.35 &     115.986209\\
$e_3$ & 0.262                  & 0.013 &    0.263156\\
$\tau_3$ & 121.21 (Days)       & 0.83 &     121.192538\\
$\varpi_3$ & 5.195 (rad)       & 0.075 &    5.175702\\
$T_0$ & 66.4219 (Days)         & 0.0016 &   66.422127\\
$v_\text{offset}$ & -560 (m/s) & 240 &      -612.095406\\ 
$R_1 (= R_2)$ & 0.28 ($R_\odot$) & 0.05 & --$^\ddag$ \\
\hline
\end{tabular}
\end{center}
* This quantity was held fixed during the dynamical analysis and the stated error comes from a separate analysis of the stellar spectrum.\\
$\dag$ This is the formal uncertainty from the MCMC analysis.  The true uncertainty would be much larger due to the uncertainty in the mass of the tertiary.\\
$\ddag$ This quantity is not part of the dynamical model.
\end{table}

\section{Discussion}\label{discussion}

Eclipse time measurements and their counterparts in the field of transiting exoplanets provide very precise measurements of the orbital phase of the various bodies.  Consequently, eclipse and transit times can be used to make similarly precise measurements of the various mass ratios and orbital parameters in multi-object systems.  In many cases the values of some parameters derived from timing measurements are significantly more precise than corresponding values from radial velocity measurements (the orbital period, for example).  Similarly, some parameters are more difficult to determine from timing measurements---depending upon the orbital configuration---such as the argument of pericentre $\varpi_3$.

Regardless, with the high precision photometry enabled by the \kepler\ spacecraft, dynamical studies of multi-object systems through timing variations has proven extremely useful as a tool to measure the orbital properties of these systems and the masses of the objects within them.  Striking examples include the planetary systems Kepler-9 \citep{Holman:2010} and Kepler-11 \citep{Lissauer:2011a} as well the hierarchical triple star KOI 126 \citep{Carter:2011} and now KOI 928.

For both KOIs 928 and 126 the very small masses determined by the dynamical analysis provide important guidance to stellar models at the low-mass end of the main sequence.  The masses in both KOI 928 and in KOI 126 are among the smallest masses observed in binary systems.  Future investigations of multiple star systems through eclipse timing variations, especially with \kepler\ photometry are likely to yield an important sample of these systems with configurations and masses heretofor unexplored.

\section*{Appendix: Table of Eclipse Times}

\begin{table*}
\begin{minipage}{100mm}
\caption{Primary eclipse times used in the analysis of this system.  The uncertainties are in days and date is BJD -- 2454900.}
\begin{tabular}{rrrrrrr}
Epoch & Date & Uncertainty & & Epoch & Date & Uncertainty \\ \hline
0	&	68.911201	&	0.012018	&	&	58	&	358.240906	&	0.013397	\\
2	&	78.873566	&	0.014027	&	&	59	&	363.281555	&	0.008515	\\
3	&	83.888573	&	0.014472	&	&	62	&	378.241272	&	0.009688	\\
4	&	88.888039	&	0.013648	&	&	63	&	383.176941	&	0.011676	\\
7	&	103.842087	&	0.015819	&	&	64	&	388.208099	&	0.016539	\\
8	&	108.809212	&	0.015918	&	&	65	&	393.203339	&	0.012011	\\
9	&	113.810135	&	0.011952	&	&	66	&	398.176483	&	0.012884	\\
10	&	118.81118	&	0.015405	&	&	67	&	403.156799	&	0.015741	\\
11	&	123.815392	&	0.014478	&	&	69	&	413.128784	&	0.014025	\\
12	&	128.791595	&	0.011868	&	&	70	&	418.098053	&	0.016432	\\
13	&	133.801773	&	0.012774	&	&	71	&	423.12796	&	0.012527	\\
14	&	138.786362	&	0.011677	&	&	72	&	428.076874	&	0.010739	\\
15	&	143.799454	&	0.015389	&	&	77	&	453.020203	&	0.013291	\\
16	&	148.767578	&	0.013429	&	&	78	&	458.011719	&	0.017929	\\
17	&	153.754303	&	0.013067	&	&	79	&	463.012665	&	0.014529	\\
18	&	158.729309	&	0.011143	&	&	80	&	468.013245	&	0.01409	\\
20	&	168.729935	&	0.024025	&	&	81	&	473.001129	&	0.015396	\\
21	&	173.70993	&	0.014521	&	&	82	&	478.007233	&	0.010852	\\
22	&	178.697556	&	0.013952	&	&	84	&	487.984894	&	0.012072	\\
23	&	183.675522	&	0.013818	&	&	85	&	492.966827	&	0.012657	\\
25	&	193.636154	&	0.0116	&	&	87	&	502.930939	&	0.017502	\\
26	&	198.624512	&	0.011541	&	&	88	&	507.90863	&	0.01025	\\
27	&	203.617828	&	0.010471	&	&	90	&	517.898987	&	0.013851	\\
28	&	208.577942	&	0.017487	&	&	91	&	522.859192	&	0.012841	\\
30	&	218.560303	&	0.014438	&	&	92	&	527.869263	&	0.014128	\\
32	&	228.562958	&	0.010952	&	&	93	&	532.848816	&	0.010089	\\
35	&	243.556519	&	0.012086	&	&	94	&	537.822205	&	0.009151	\\
36	&	248.56221	&	0.009725	&	&	95	&	542.79834	&	0.01213	\\
37	&	253.544861	&	0.014524	&	&	96	&	547.783997	&	0.013464	\\
38	&	258.5065	&	0.013676	&	&	100	&	567.72467	&	0.010141	\\
39	&	263.502747	&	0.007326	&	&	101	&	572.750916	&	0.011728	\\
40	&	268.487061	&	0.019154	&	&	102	&	577.744812	&	0.011363	\\
41	&	273.472717	&	0.017923	&	&	103	&	582.715576	&	0.00987	\\
42	&	278.422852	&	0.011309	&	&	104	&	587.740295	&	0.00963	\\
44	&	288.437225	&	0.022191	&	&	105	&	592.697571	&	0.011524	\\
45	&	293.40271	&	0.012995	&	&	106	&	597.724915	&	0.016151	\\
48	&	308.357758	&	0.011893	&	&	107	&	602.669006	&	0.010604	\\
49	&	313.352875	&	0.0121	&	&	108	&	607.686768	&	0.016118	\\
50	&	318.340729	&	0.011106	&	&	109	&	612.672607	&	0.020504	\\
51	&	323.327301	&	0.008211	&	&	111	&	622.648499	&	0.011646	\\
52	&	328.286224	&	0.014265	&	&	112	&	627.644287	&	0.020589	\\
54	&	338.289398	&	0.013852	&	&	113	&	632.61499	&	0.020151	\\
55	&	343.273407	&	0.017753	&	&	114	&	637.621765	&	0.014057	\\
56	&	348.262726	&	0.011482	&	&	115	&	642.560303	&	0.016378	\\
57	&	353.280609	&	0.013004	&	&	116	&	647.573486	&	0.018376	\\
\end{tabular}
\end{minipage}
\end{table*}

\begin{table*}
\begin{minipage}{100mm}
\caption{Secondary eclipse times used in the analysis of this system.  The uncertainties are in days and date is BJD -- 2454900.}
\begin{tabular}{rrrrrrr}
Epoch & Date & Uncertainty & & Epoch & Date & Uncertainty \\ \hline
0.5	&	71.423409	&	0.006498	&	&	58.5	&	360.782715	&	0.007054	\\
1.5	&	76.392502	&	0.00588	&	&	59.5	&	365.763611	&	0.008296	\\
2.5	&	81.402039	&	0.008043	&	&	60.5	&	370.747498	&	0.012001	\\
3.5	&	86.383942	&	0.008512	&	&	62.5	&	380.709595	&	0.007119	\\
5.5	&	96.323235	&	0.009293	&	&	63.5	&	385.715057	&	0.008714	\\
7.5	&	106.326385	&	0.007729	&	&	64.5	&	390.701111	&	0.005317	\\
8.5	&	111.310364	&	0.006473	&	&	65.5	&	395.675171	&	0.007651	\\
10.5	&	121.309601	&	0.004985	&	&	66.5	&	400.682617	&	0.009007	\\
11.5	&	126.311958	&	0.006765	&	&	67.5	&	405.644043	&	0.007943	\\
12.5	&	131.331589	&	0.01102	&	&	68.5	&	410.630402	&	0.006404	\\
13.5	&	136.303024	&	0.006376	&	&	69.5	&	415.617615	&	0.006804	\\
15.5	&	146.279373	&	0.00769	&	&	70.5	&	420.602356	&	0.006952	\\
17.5	&	156.270187	&	0.006814	&	&	71.5	&	425.577698	&	0.006294	\\
18.5	&	161.239426	&	0.007947	&	&	72.5	&	430.564819	&	0.007979	\\
19.5	&	166.216278	&	0.007817	&	&	74.5	&	440.536743	&	0.007105	\\
20.5	&	171.195129	&	0.010561	&	&	75.5	&	445.524628	&	0.006919	\\
21.5	&	176.193253	&	0.008447	&	&	76.5	&	450.513214	&	0.006315	\\
22.5	&	181.179413	&	0.007294	&	&	77.5	&	455.510071	&	0.009652	\\
23.5	&	186.157288	&	0.006237	&	&	79.5	&	465.511963	&	0.005533	\\
24.5	&	191.150574	&	0.006963	&	&	80.5	&	470.492767	&	0.007039	\\
25.5	&	196.111954	&	0.007402	&	&	81.5	&	475.479889	&	0.006334	\\
26.5	&	201.097565	&	0.00778	&	&	82.5	&	480.49292	&	0.00597	\\
27.5	&	206.10289	&	0.005941	&	&	83.5	&	485.493591	&	0.007924	\\
29.5	&	216.075058	&	0.007029	&	&	87.5	&	505.414032	&	0.007591	\\
30.5	&	221.070267	&	0.006524	&	&	89.5	&	515.38208	&	0.008315	\\
31.5	&	226.0793	&	0.011133	&	&	90.5	&	520.385315	&	0.010596	\\
32.5	&	231.038559	&	0.006171	&	&	91.5	&	525.372253	&	0.011837	\\
34.5	&	241.022278	&	0.006232	&	&	93.5	&	535.328247	&	0.007012	\\
35.5	&	246.033356	&	0.007357	&	&	94.5	&	540.318115	&	0.006934	\\
36.5	&	251.029495	&	0.010043	&	&	95.5	&	545.295166	&	0.012277	\\
39.5	&	265.999756	&	0.006732	&	&	98.5	&	560.269043	&	0.008803	\\
40.5	&	270.991089	&	0.009914	&	&	99.5	&	565.269775	&	0.017692	\\
41.5	&	275.947021	&	0.008132	&	&	100.5	&	570.237183	&	0.02373	\\
43.5	&	285.930481	&	0.006291	&	&	102.5	&	580.230408	&	0.013571	\\
44.5	&	290.914825	&	0.008349	&	&	103.5	&	585.216187	&	0.01391	\\
45.5	&	295.908386	&	0.006239	&	&	104.5	&	590.21814	&	0.00964	\\
46.5	&	300.888611	&	0.006748	&	&	105.5	&	595.221619	&	0.009139	\\
47.5	&	305.86322	&	0.005782	&	&	107.5	&	605.207397	&	0.01174	\\
48.5	&	310.847107	&	0.00527	&	&	108.5	&	610.196777	&	0.013152	\\
49.5	&	315.847626	&	0.006766	&	&	109.5	&	615.191162	&	0.016818	\\
50.5	&	320.815552	&	0.006921	&	&	110.5	&	620.16748	&	0.016078	\\
51.5	&	325.818909	&	0.00729	&	&	111.5	&	625.14502	&	0.007044	\\
53.5	&	335.803284	&	0.00605	&	&	112.5	&	630.137756	&	0.009316	\\
54.5	&	340.782318	&	0.006592	&	&	113.5	&	635.125061	&	0.008281	\\
55.5	&	345.763702	&	0.005504	&	&	114.5	&	640.097046	&	0.019446	\\
56.5	&	350.77002	&	0.004739	&	&	115.5	&	645.087463	&	0.007803	\\
57.5	&	355.767761	&	0.007724	&	&	-	&	-	&	-	\\
\end{tabular}
\end{minipage}
\end{table*}

\section*{Acknowledgments}

\kepler\ is NASA's tenth Discovery mission with funding provided by NASA's Science Mission Directorate.  D. C. F. acknowledge NASA support through Hubble Fellowship grants 
\#HF-51272.01-A and \#HF-51267.01-A, awarded by STScI and operated by AURA under contract NAS 5-26555.

\bsp

\label{lastpage}

\end{document}